# DO WE NEED (OR WANT) A BOSONIC GLUE TO PAIR ELECTRONS IN HIGH $T_C$ SUPERCONDUCTORS?

Abstract: Many investigators have joined the search for a bosonic "glue" which is hypothecated to be the mechanism which binds the electron pairs in the cuprate high Tc superconductors, often referring to the "Eliashberg formalism" which was developed to reveal the role of phonons in the conventional polyelectronic metal superconductors. In this paper we point out that the picture of boson exchange is a "folklore" description of the pairing process with no rigorous basis. The problem of pairing is always that of evading the strong Coulomb vertex, the repulsive core of the interaction; we discuss the different means by which the two types of superconductors accomplish this feat.

For many years papers have been appearing which discuss the high Tc superconductors in parallel with the conventional metallic superconductors, as involving electron pairs bound together by exchange of some bosonic excitation, analogously to the presumed role of phonons in the ordinary metals. Many alternatives have been proposed for this bosonic "glue": the optical or acoustic phonon spectrum[1,2], the 41 mev magnetic resonance mode[3], a mysterious broad bosonic spectrum responsible for the anomalous resistivity[4,5] an equally mysterious mode responsible for the "dip" in the ARPES spectrum[6], and "antiferromagnetic spin fluctuations"[7], are only a few of the proposals which have been put forward. (I make no attempt to give a complete citation list, concentrating on papers which have garnered recent attention.) Common to all such papers is that the pairing must be mediated by some mode which has dynamic character, rather than being caused by a simple attractive vertex scattering the electron pair in a singlet channel.

I argue here that this need for a bosonic "glue" is of the nature of folklore rather than of scientific logic; it comes from the entirely inappropriate assumption that superconductivity in these materials is described by some equivalent "Eliashberg" formalism, a formalism (developed by Morel and Anderson[8] and Schrieffer et al[9] from Eliashberg's treatment of the crude Frohlich Hamiltonian for the electron-phonon system) which is appropriate only to describe the "dynamic screening" mechanism which explains superconductivity in polyelectronic metals.

Let us first explain the dynamic screening mechanism. We must realize that electrons only interact, to a very good approximation, via the Coulomb interaction, and the bare Coulomb interaction is strongly repulsive. There are two basic ways to avoid this and bind electrons in pairs nonetheless: via screening or via the mechanism suggested by Pitaevskii[10] and by Bruckner et al[11] of making a pair state orthogonal to the repulsive core of the Coulomb interaction, such as a d-wave. The first is used in conventional superconductors, the second in the high Tc cuprates and many other unconventional superconductors.

In the former case we may write the interaction as

$$V(q,\omega) = e^2/q^2\varepsilon(q,\omega) \qquad (1)$$

where $\varepsilon$ is the momentum and frequency-dependent dielectric constant. Screening by the plasma modes of the electron gas reduces the high-frequency interaction to the screened Coulomb form in which the denominator becomes $(q^2+k_s^2)$.

The low-frequency interaction is further screened by the phonons, which may be introduced either as an additional component of the dielectric constant due to phonons or via a direct electron-phonon interaction, as in the Eliashberg formalism. The Eliashberg formalism neglects all vertex corrections on the basis that the phonon frequencies are low. But the former description more completely clarifies the physics, and in particular brings out the limitations on the magnitude of the interaction, in that $\varepsilon$ must remain positive at zero frequency to maintain stability of the lattice (give or take minor local field corrections), so that the attractive phonon interaction—which in reference 9 we summarized with a dimensionless parameter $\lambda$--may never be much bigger, and is normally smaller, than the screened Coulomb repulsion, which we summarized with a parameter $\mu$. The net interaction is repulsive even in the phonon case, and it is only its dynamic structure which allows it to bind pairs.

How then do we ever get bound pairs, if the interaction is never attractive? This occurs by taking advantage of the difference in frequency scales of the two pieces of the interaction. There is a ladder-sum renormalization, or pseudization, of the Coulomb repulsion to a smaller value

$$\mu^* = \frac{\mu}{1 + \mu \ln(E_F/\omega_D)} \quad (2)$$

so that the effective interaction is then $-(\lambda - \mu^*) < 0$. This combination is what appears in the Dynes-Macmillan semi-empirical formula for Tc in these conventional superconductors. Thus one may say that superconductivity results from the bosonic interaction via phonons; but it is equally valid to say instead that it results from the ladder renormalization which gives us $\mu^*$ rather than $\mu$, and which does not show up in an Eliashberg analysis—it is a vertex correction to the screened Coulomb interaction.

The above is an instructive example to show that the "Eliashberg" theory is by no means a formalism which universally demonstrates the nature of the pairing interaction; it is merely a convenient effective theory of any portion of the interactions which comes from low-frequency bosons. There is no reason to believe that this framework is appropriate to describe a system where the pairing depends on entirely different physics.

Such a system is the $CuO_2$ planes of the cuprate superconductors. The key difference from polyelectronic metals is that the relevant electrons are in a single band, the antibonding $d_{x2-y2}$-$p_\sigma$ band which may be built up from a Wannier function of $x^2$-$y^2$ symmetry, with a spectrum bounded both at high and low energies. In such a band the ladder-sum renormalization of the local Coulomb repulsion, leading to the pseudopotential µ*, simply does not work: the band doesn't contain enough of Hilbert space to remove the local part of the repulsion. (To describe this difficulty technically, since there is only one Wannier function per atom, the shape of the function representing two electrons on the same atom can't be modified, which is what happens when we form the pseudopotential.) This is why the Hubbard repulsion U is all-important in this band, a fact which is confirmed by the Mott insulator character of the undoped cuprate, but its effects are not at all confined to the low-doped system. In low-energy wave functions the electrons avoid being on the same site by means of a Jastrow factor which is conveniently modeled by the Gutzwiller procedure. As a consequence they scatter each other very strongly[12] and most of the broad structure in their Green's functions is caused by U. This structure may naively be ascribed to coupling to a broad spectrum of bosonic modes[13] but has nothing to do with pair binding.

A second consequence of U is the appearance, perturbatively in 1/U, of a large antiferromagnetic exchange coupling attracting electrons of opposite spins on neighboring sites. The matrix elements responsible for this connect states of very high energies and the corresponding interaction vertex has only high-frequency dynamics, so it is unrelated to a "glue". There is a common misapprehension that it has some relation to low-frequency spin fluctuations[14] but that is incorrect.

In order to avoid this repulsive potential these systems use the alternative Pitaevskii-Brueckner-Anderson scheme with pairing orthogonal to the local potential. Two such pairings exist, d-wave and "extended s-wave", and both are used, but only one appears as a superconducting gap.[15] Because of the large magnitude of J, of order >1000K, the pairing is very strong, but only a small fraction of this pairing energy shows up as a superconducting Tc, for various rather complicated reasons.

The crucial point made in this discussion is that there are two very strong interactions that we know, both a priori and because of incontrovertible experimental evidence, are present in the cuprates. Neither is properly described by a bosonic "glue"—a "bubble" sum in diagram terms-- and between the two it is easy to account for the existence of antiferromagnetism, d-wave superconductivity, and many other phenomena of high Tc superconductiv Whether any additional "glue" exists is of lesser interest. We have a mammoth and an elephant in our refrigerator—do we care much if there is also a mouse?

I acknowledge extensive recent discussions with J C Davis, T Timusk, L Pietronero, N P Ong, and Ali Yazdani.